\documentclass{elsart}

\usepackage[square,comma]{natbib}
\usepackage{graphicx}
\usepackage{pxfonts}
\usepackage{lineno}

\usepackage{amssymb}

\journal{}

\begin{document}

\thispagestyle{empty}
\begin{Large}
\textbf{DEUTSCHES ELEKTRONEN-SYNCHROTRON}

\textbf{\large{Ein Forschungszentrum der Helmholtz-Gemeinschaft}\\}
\end{Large}

DESY 11-083

May 2011

\begin{eqnarray}
\nonumber &&\cr \nonumber && \cr \nonumber &&\cr
\end{eqnarray}
\begin{eqnarray}
\nonumber
\end{eqnarray}
\begin{center}
\begin{Large}
\textbf{Microbunch preserving in-line system for an APPLE II helical
radiator at the LCLS baseline }
\end{Large}
\begin{eqnarray}
\nonumber &&\cr \nonumber && \cr
\end{eqnarray}

\begin{large}
Gianluca Geloni,
\end{large}
\textsl{\\European XFEL GmbH, Hamburg}
\begin{large}

Vitali Kocharyan and Evgeni Saldin
\end{large}
\textsl{\\Deutsches Elektronen-Synchrotron DESY, Hamburg}
\begin{eqnarray}
\nonumber
\end{eqnarray}
\begin{eqnarray}
\nonumber
\end{eqnarray}
ISSN 0418-9833
\begin{eqnarray}
\nonumber
\end{eqnarray}
\begin{large}
\textbf{NOTKESTRASSE 85 - 22607 HAMBURG}
\end{large}
\end{center}
\clearpage
\newpage

\begin{frontmatter}



\title{Microbunch preserving in-line system for an APPLE II helical radiator at
the LCLS baseline }


\author[XFEL]{Gianluca Geloni\thanksref{corr},}
\thanks[corr]{Corresponding Author. E-mail address: gianluca.geloni@xfel.eu}
\author[DESY]{Vitali Kocharyan}
\author[DESY]{and Evgeni Saldin}

\address[XFEL]{European XFEL GmbH, Hamburg, Germany}
\address[DESY]{Deutsches Elektronen-Synchrotron (DESY), Hamburg,
Germany}

\begin{abstract}
In a previous work we proposed a scheme for polarization control at
the LCLS baseline, which exploited the microbunching from the planar
undulator. After the baseline undulator, the electron beam is
transported through a drift by a FODO focusing system, and through a
short helical radiator. The microbunching structure can be
preserved, and intense coherent radiation is emitted in the helical
undulator at fundamental harmonic.  The driving idea of this
proposal is that the background linearly-polarized radiation from
the baseline undulator is suppressed by spatial filtering. Filtering
is achieved by letting radiation and electron beam through Be slits
upstream of the helical radiator, where the radiation spot size is
about ten times larger than the electron beam transverse size.
Several changes considered in the present paper were made to improve
the previous design. Slits are now placed immediately behind the
helical radiator. The advantage is that the electron beam can be
spoiled by the slits, and narrower slits width can be used for
spatial filtering. Due to this fundamental reason, the present setup
is shorter than the previous one. The helical radiator is now placed
immediately behind the SHAB undulator. It is thus sufficient to use
the existing FODO focusing system of the SHAB undulator for
transporting the modulated electron beam. This paper presents
complete GENESIS code calculations for the new design, starting from
the baseline undulator entrance up to the helical radiator exit
including the modulated electron beam transport by the SHAB FODO
focusing system.
\end{abstract}

\end{frontmatter}



\section{\label{sec:uno}  Introduction}

The LCLS baseline setup includes a planar undulator which produces
intense linearly-polarized light in the wavelength range between
$0.12$ nm and $2.2$ nm \cite{LCLS2}. There is, however, an
increasing demand for circularly-polarized X-ray pulses, especially
in the soft X-ray region and, in particular, in the spectral range
between $550$ eV and $900$ eV, which covers many important
absorption edges of transition metals like Cr, Mn, Fe, Co, Ni. The
relevance of these metals is evident if one reminds, for example,
that the elementary ferromagnetic F, Co, Ni, form a basis for
information storage. The LCLS covers the photon energy range down to
$550$ eV, so that the region between $550$ eV and $900$ eV can now
be covered by the LCLS baseline in the fundamental harmonic.

Several schemes using helical undulators have been proposed for
polarization control at the LCLS \cite{GENG,KUSK,OURC}. Since the
SASE process already provides electron beam  microbunching,  the
microbunches radiates coherently when passing through an helical
undulator tuned at the same radiation wavelength. Therefore, it is
not necessary that all the undulators in the line be helical. Along
these lines of reasoning, all proposed schemes \cite{GENG,KUSK,OURC}
exploit the microbunching of the planar undulator, and make use of a
short helical radiator at the end of the undulator beamline.
However, the exploitation of planar undulator leads to background
problem, since the linearly-polarized radiation from the baseline
undulator should be suppressed.

\begin{figure}[tb]
\begin{center}
\includegraphics[width=0.75\textwidth]{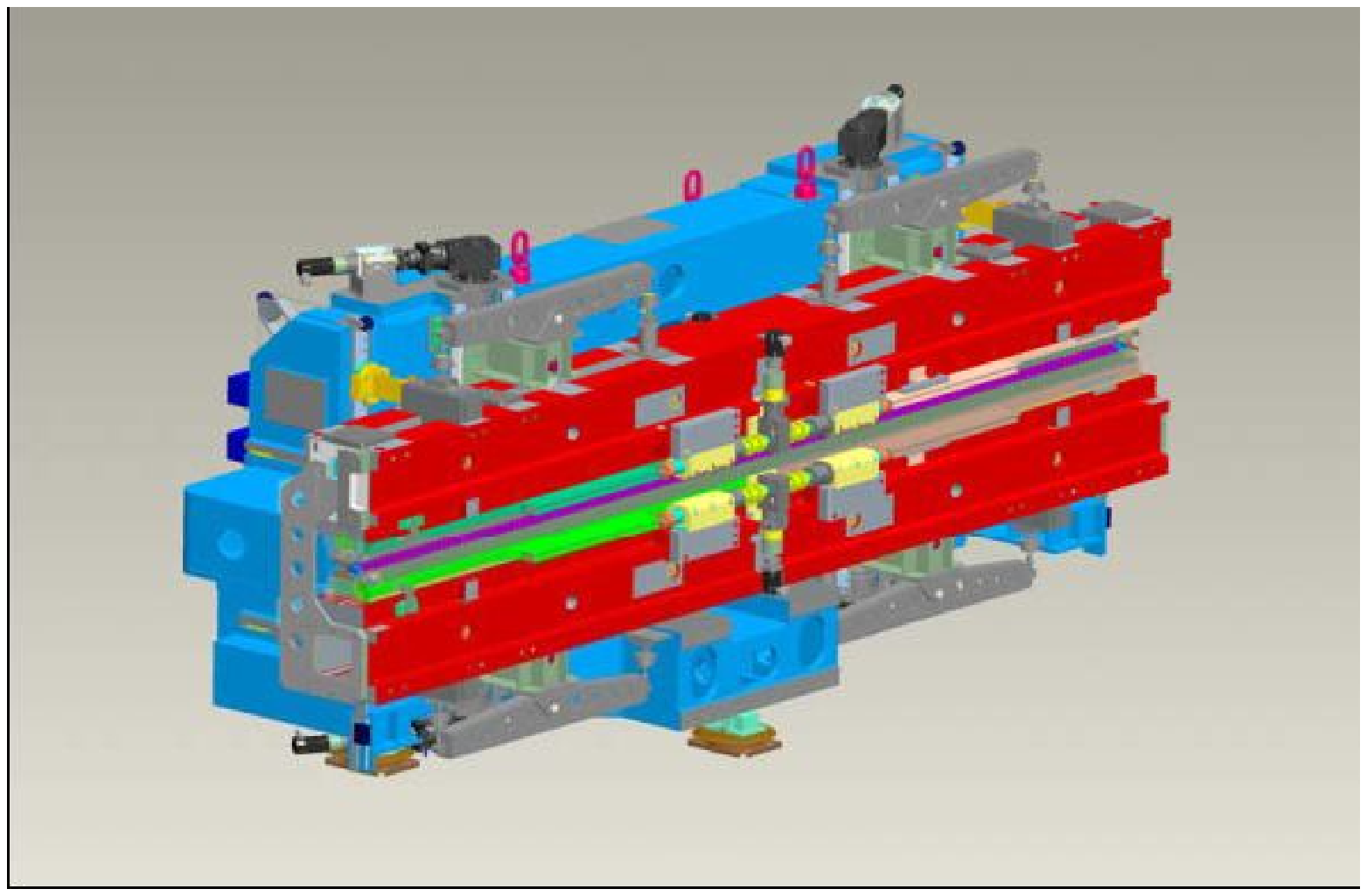}
\end{center}
\caption{Mechanical layout of the APPLE II undulator module
(Courtesy of M. Tischer).} \label{AppleII}
\end{figure}
It has been proposed \cite{GENG} that the radiation in the helical
radiator can be tuned to the second harmonic (second harmonic
afterburner helical radiator), and is therefore characterized by a
different frequency compared to the linearly polarized radiation,
tuned at the fundamental. However, for the LCLS this option can be
extended only down to 1 keV and cannot cover the most interesting
region between $550$ eV and $900$ eV. Another possible solution
\cite{KUSK}is based on the use of APPLE III type undulator modules.
At the LCLS, saturation of the linearly polarized radiation at $1.5$
nm is reached after $6$ undulator modules, with a saturation power
is about 10 GW. In order to reduce the linearly-polarized
background, and to reach a degree of circular polarization larger
than $95 \%$ , APPLE III undulators need to be installed before the
linearly polarized output reaches the $0.1$ GW power level, i.e. one
needs to install three undulator modules. The main drawback of this
scheme is constituted by the technical challenge of producing long
helical insertion devices, since APPLE III type undulators have not
yet come into operation. Finally, in \cite{OURC} we proposed a third
option which mainly consists of sending the electron beam, after the
passage through the baseline undulator, through a $40$ m -long
straight section, and subsequently through a short helical radiator.
The background radiation from the baseline undulator is suppressed
by letting radiation and electron beam through Beryllium vertical
and horizontal slits upstream the helical radiator, where the
radiation spot size is about ten times larger than the electron
bunch transverse size.  The option proposed in \cite{OURC} has
advantages in terms of cost and time, not only because helical
undulator would be only $5$ m-long, but also because we can afford
to use the existing design of APPLE II type undulators, Fig.
\ref{AppleII}, improved for PETRA III \cite{BAHR}.

This paper proposes several modifications to improve the design
presented in \cite{OURC}. First, it seems reasonable that slits are
placed immediately behind the helical radiator. The advantage of
this choice over the previous one \cite{OURC} is that the quality of
the electron beam can now be spoiled by slits, since they are placed
after helical radiator, and therefore a narrower width of the slits
can be used for spatial filtering. In this way, the overall setup is
fundamentally shorter than the previous one. Second, the helical
radiator now is placed immediately behind the second harmonic
afterburner (SHAB) undulator. We demonstrated by simulations
performed with the code GENESIS \cite{GENE} that in order to
transport the microbunched electron beam through the $20$ m-ling
straight section corresponding to the five SHAB undulator modules,
it is sufficient to use, as a focusing lattice, the existing SHAB
undulator FODO system with usual $10$ m betatron function function.

The new setup proposed in this work is extremely compact, and is
composed of as few as two elements: the $5$ m-long APPLE II
undulator module and the slits. The cost for a single APPLE II
undulator module amounts to about $1.5$ million dollars, and
manufacturing time can be estimated in two years. Altogether, in
this paper we offer a new option for polarization control at the
LCLS baseline which promises excellent cost-effective and risk-free
results.

\section{\label{sec:due} Circular polarization control scheme with spatial filtering out the
linearly-polarized radiation behind the helical radiator. }

\begin{figure}[tb]
\includegraphics[width=1.0\textwidth]{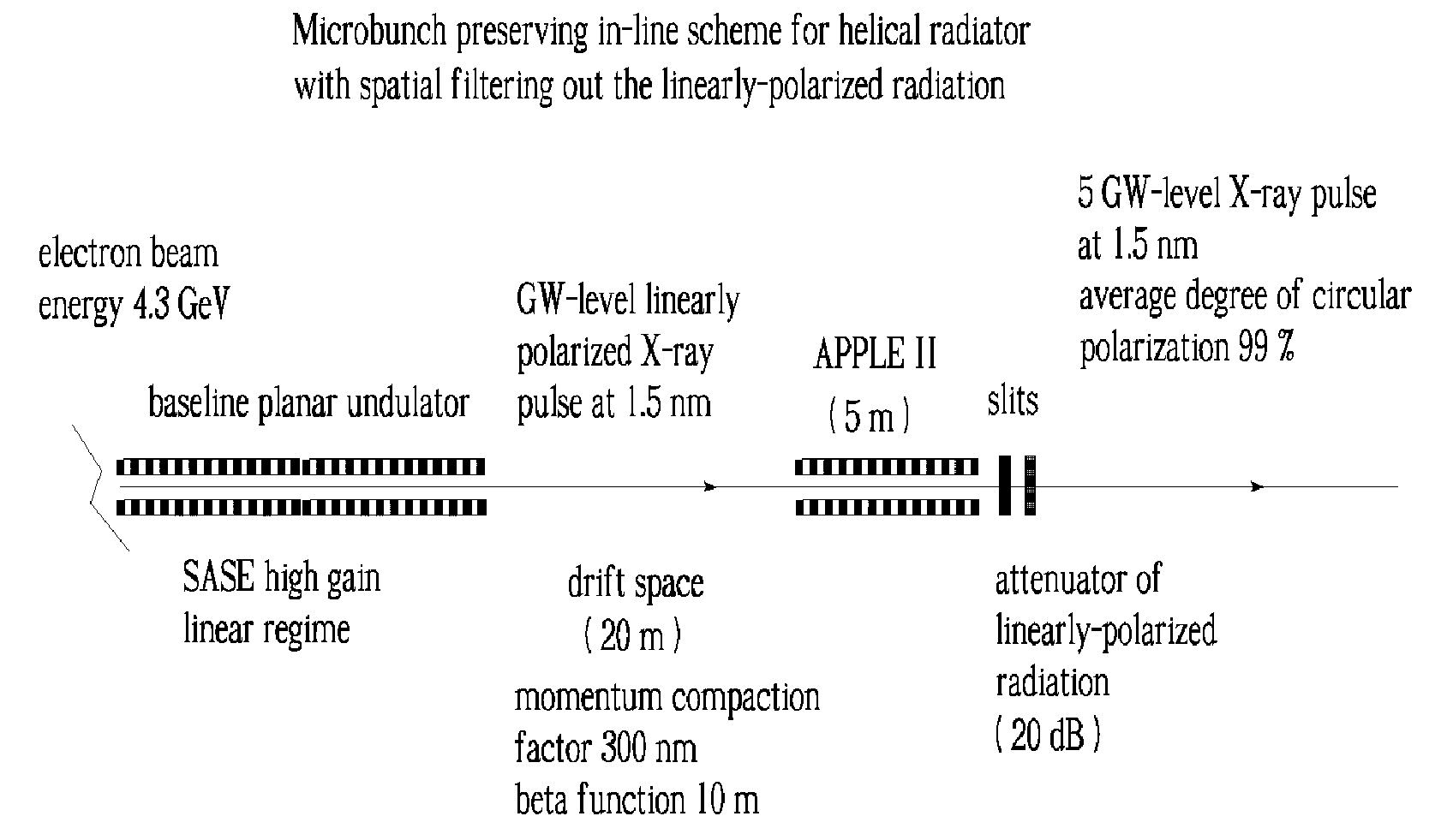}
\caption{Concept of circular polarization control at the LCLS
baseline. After the baseline undulator, the electron beam is
propagated along a $20$ m-long straight section and subsequently
passes through a helical radiator. The microbunching is preserved,
and intense coherent radiation is emitted in the helical radiator.
Linearly-polarized radiation from the baseline undulator is easily
suppressed by spatial filtering with the help of slits downstream
the helical radiator, which do not attenuate the
circularly-polarized radiation from the radiator. } \label{beta3}
\end{figure}

\begin{figure}[tb]
\includegraphics[width=1.0\textwidth]{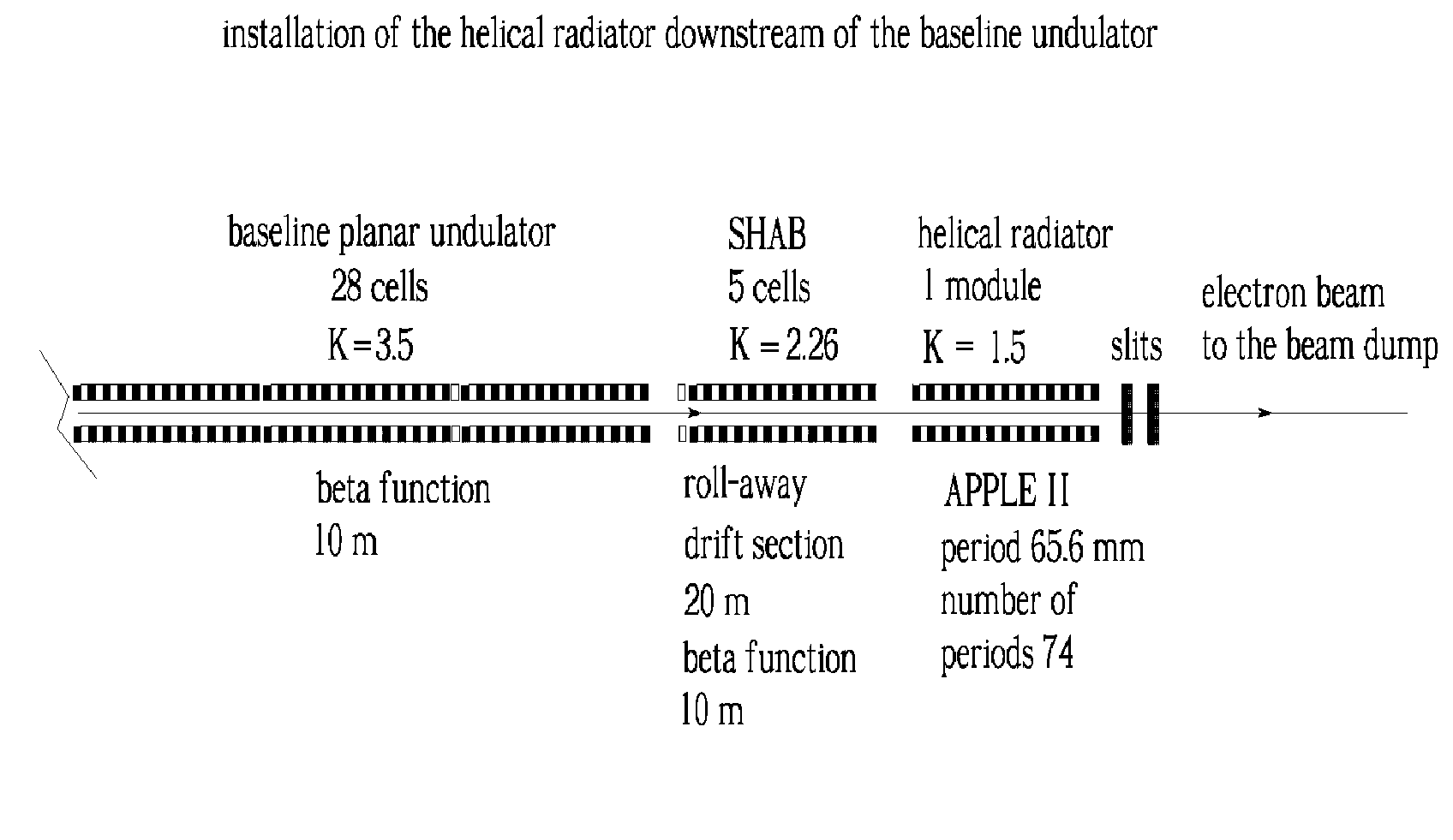}
\caption{The installation of helical radiator and slits downstream
the LCLS baseline undulator will allow to produce high-power, highly
circularly-polarized soft X-ray radiation.} \label{beta1}
\end{figure}

\begin{figure}[tb]
\includegraphics[width=1.0\textwidth]{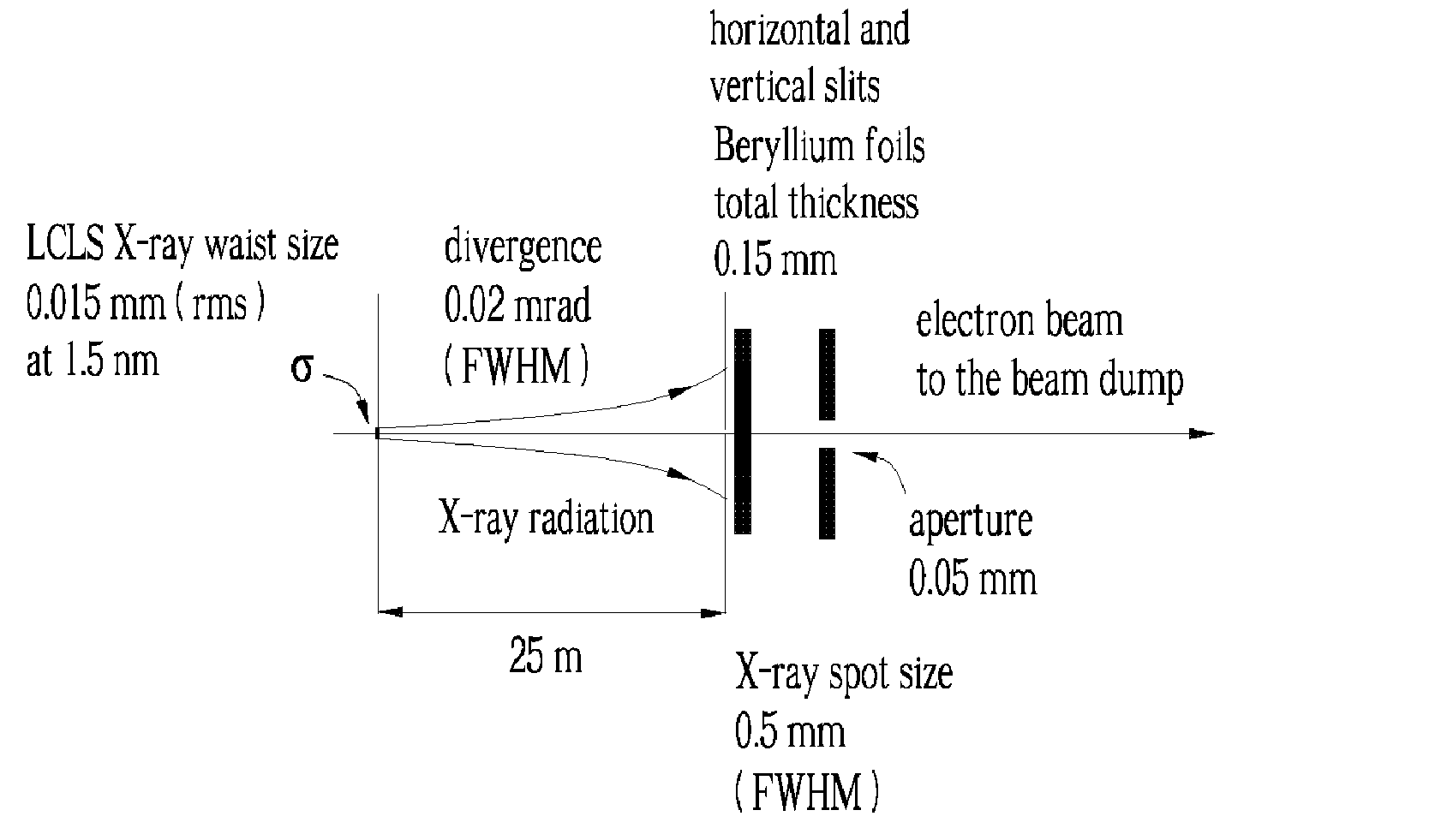}
\caption{Simple method for suppressing the linearly-polarized soft
X-ray radiation from the LCLS baseline undulator. The
linearly-polarized background can be eliminated by using a spatial
window positioned downstream of the helical radiator exit. This can
be practically implemented by letting radiation and electron bunch
through vertical and horizontal slits positioned $25$ m downstream
of the planar undulator, where the linearly-polarized radiation
pulse is characterized by a tenfold larger spot size compared with
the circularly-polarized radiation spot size.} \label{beta2}
\end{figure}
The principle upon which our scheme for polarization control is
based is straightforward, and is illustrated in Fig. \ref{beta3},
Fig. \ref{beta1} and Fig. \ref{beta2}. The electron beam first goes
through the baseline undulator, it produces SASE radiation and is
modulated in energy and density.

As in \cite{OURC}, we assume here as well that the five second
harmonic afterburner (SHAB) modules are rolled away \cite{BAIL} from
the beam line. In this way we provide a total $20$ m-long straight
section for the electron beam transport, corresponding to the length
of the SHAB modules. At the end of the straight section, that is
immediately behind the SHAB undulator, we install a $5$ m-long APPLE
II type undulator. While passing through this helical radiator, the
microbunched electron beam produces intense bursts of radiation in
any selected state of polarization. Subsequently, the polarized
radiation from the APPLE II undulator, the linearly polarized
radiation from the baseline undulator and the electron beam pass
through horizontal and vertical slits. This results in a suppression
of the linearly polarized radiation. In fact, since the slits are
positioned $25$ m downstream of the planar undulator, the
linearly-polarized radiation has about ten times larger spot size
compared to the circularly-polarized radiation spot size, and the
background radiation power can therefore be diminished of two orders
of magnitude. The slits can be made out of Beryllium foils, for a
total thickness of about $0.15$ mm.  Such foils block the radiation,
but lets the electrons go through.

In order to understand the effect of the foils on the electrons, we
need to address multiple Coulomb scattering in the foils. We obtain
a spoiled normalized emittance $\epsilon_\mathrm{n} \simeq 20
~\mu$m. This normalized emittance is well within the acceptance of
the beamline optics. The advantage of the spoiling scheme is that
radiation is attenuated of $20$ dB, but the spoiled electron bunch
is allowed to propagate through the straight line up to the beam
dump without electron losses \cite{PREM}.

The influence of the propagation of the electron beam through the
drift section on the electron beam microbunching should be accounted
for. One should account for the fact that straight section acts as a
dispersive element with momentum compaction factor $R_{56} \simeq
300$ nm at an electron beam energy of $4.3$ GeV. The influence of
the betatron motion should be further accounted for. In fact, the
finite angular divergence of the electron beam, which is linked with
the betatron function, leads to longitudinal velocity spread
yielding microbunching suppression. In the next Section we present a
comprehensive study of these effects. We simulated the evolution of
the microbunching along the straight section, and we concluded that
the transport of the microbunched electron beam through the $20$ m -
long straight section does not constitute a serious problem for the
realization of the proposed scheme.

\section{\label{sec:tre} FEL simulations}

In this Section we report on a feasibility study performed with the
help of the FEL code GENESIS 1.3 \cite{GENE} running on a parallel
machine. We will present a feasibility study for our method of
polarization control at the LCLS, based on a statistical analysis
consisting of $100$ runs. Parameters used in the simulations for the
low-charge mode of operation are presented in Table \ref{tt1}. The
choice of the low-charge mode of operation is motivated by
simplicity.

\begin{table}
\caption{Parameters for the low-charge mode of operation at LCLS
used in this paper.}

\begin{small}\begin{tabular}{ l c c}
\hline & ~ Units &  ~ \\ \hline
Undulator period      & mm                  & 30     \\
K parameter (rms)     & -                   & 2.466  \\
Wavelength            & nm                  & 1.5   \\
Energy                & GeV                 & 4.3   \\
Charge                & nC                  & 0.02 \\
Bunch length (rms)    & $\mu$m              & 1    \\
Normalized emittance  & mm~mrad             & 0.4    \\
Energy spread         & MeV                 & 1.5   \\
\hline
\end{tabular}\end{small}
\label{tt1}
\end{table}

\begin{figure}
\includegraphics[width=0.5\textwidth]{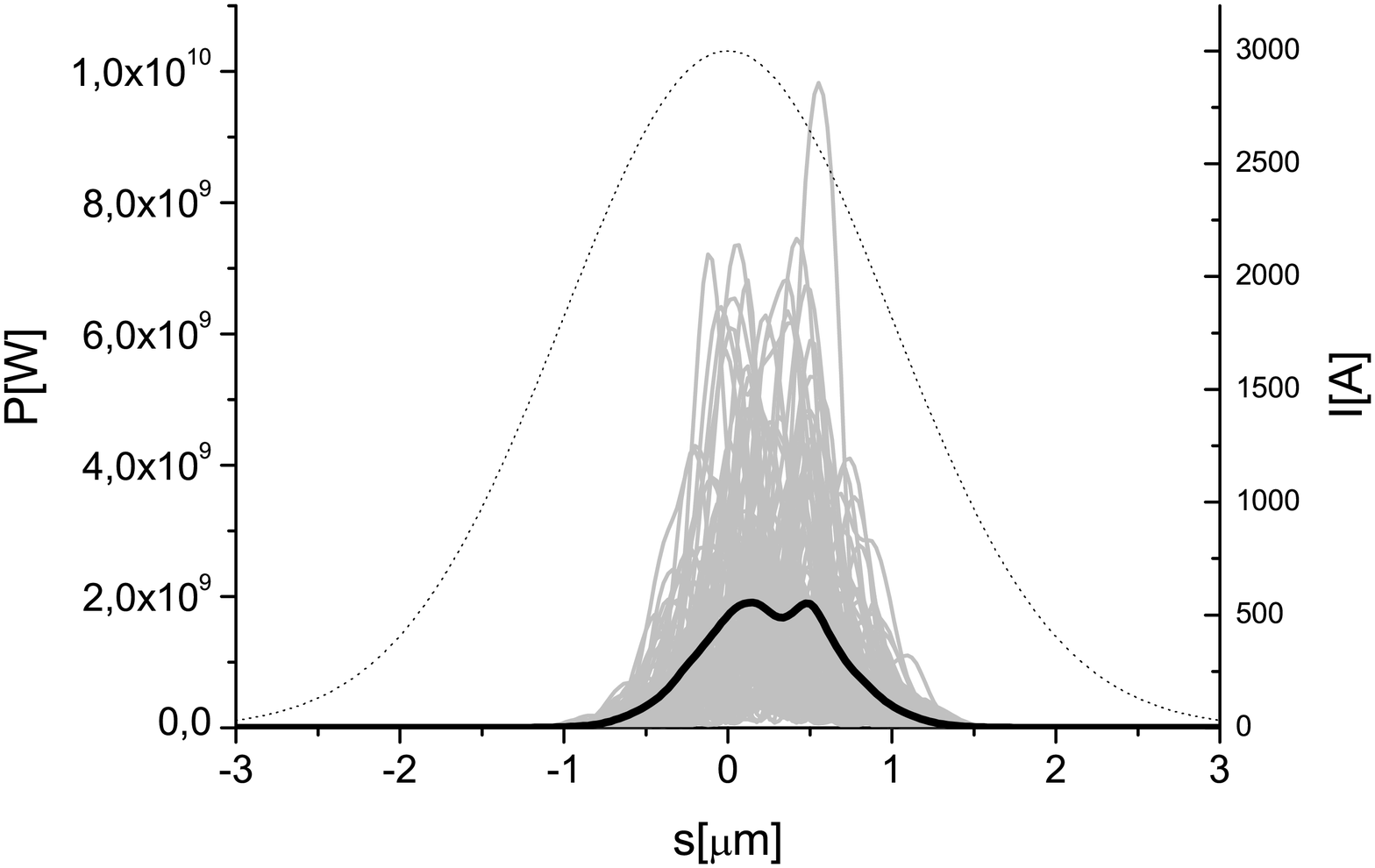}
\includegraphics[width=0.5\textwidth]{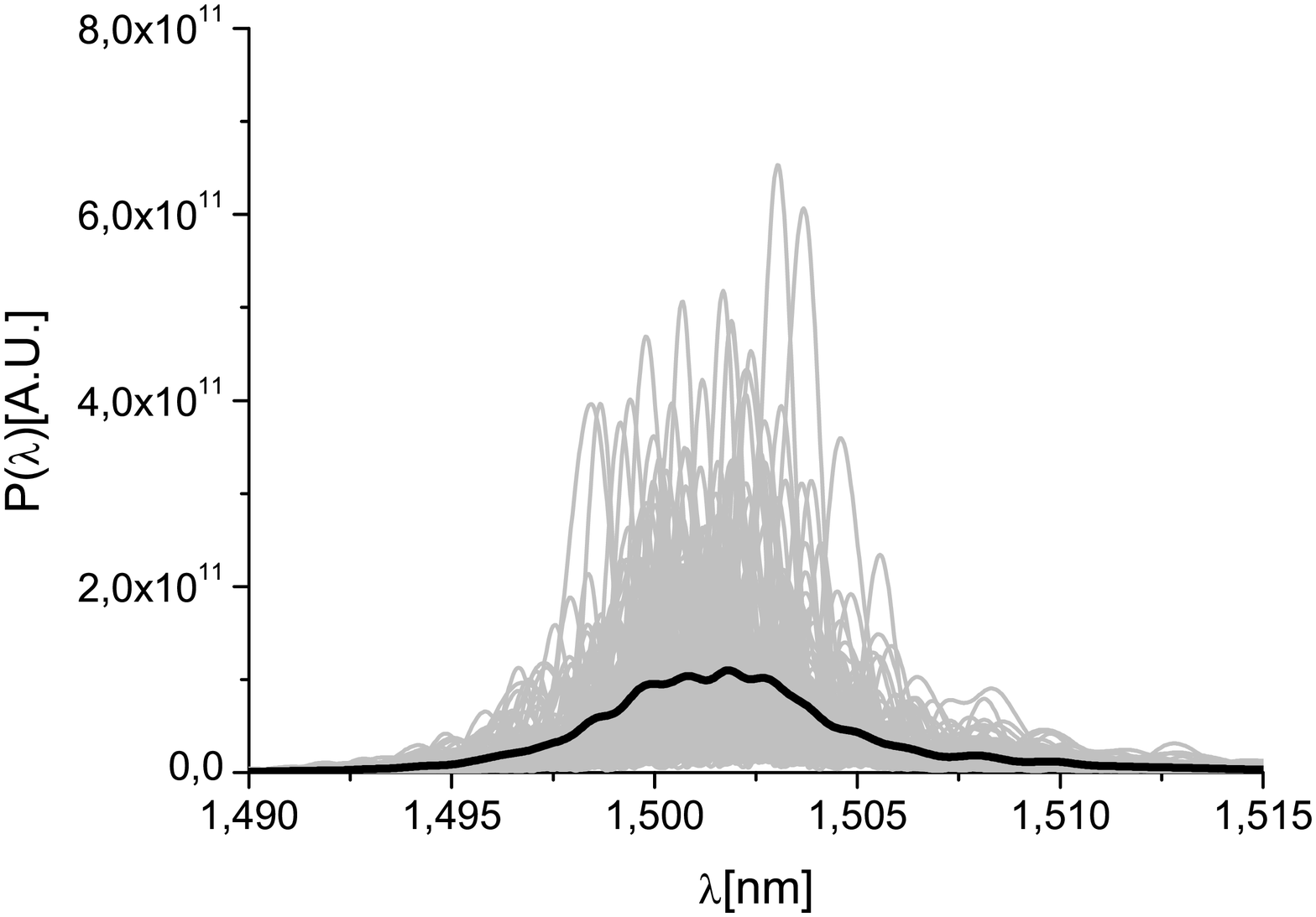}
\caption{Left plot: power distribution after the first SASE
undulator (5 cells). The dotted line refers to the original electron
bunch profile. Right plot: spectrum after the first SASE undulator
(5 cells). Grey lines refer to single shot realizations, the black
line refers to the average over a hundred realizations. }
\label{SASE}
\end{figure}

\begin{figure}
\includegraphics[width=0.5\textwidth]{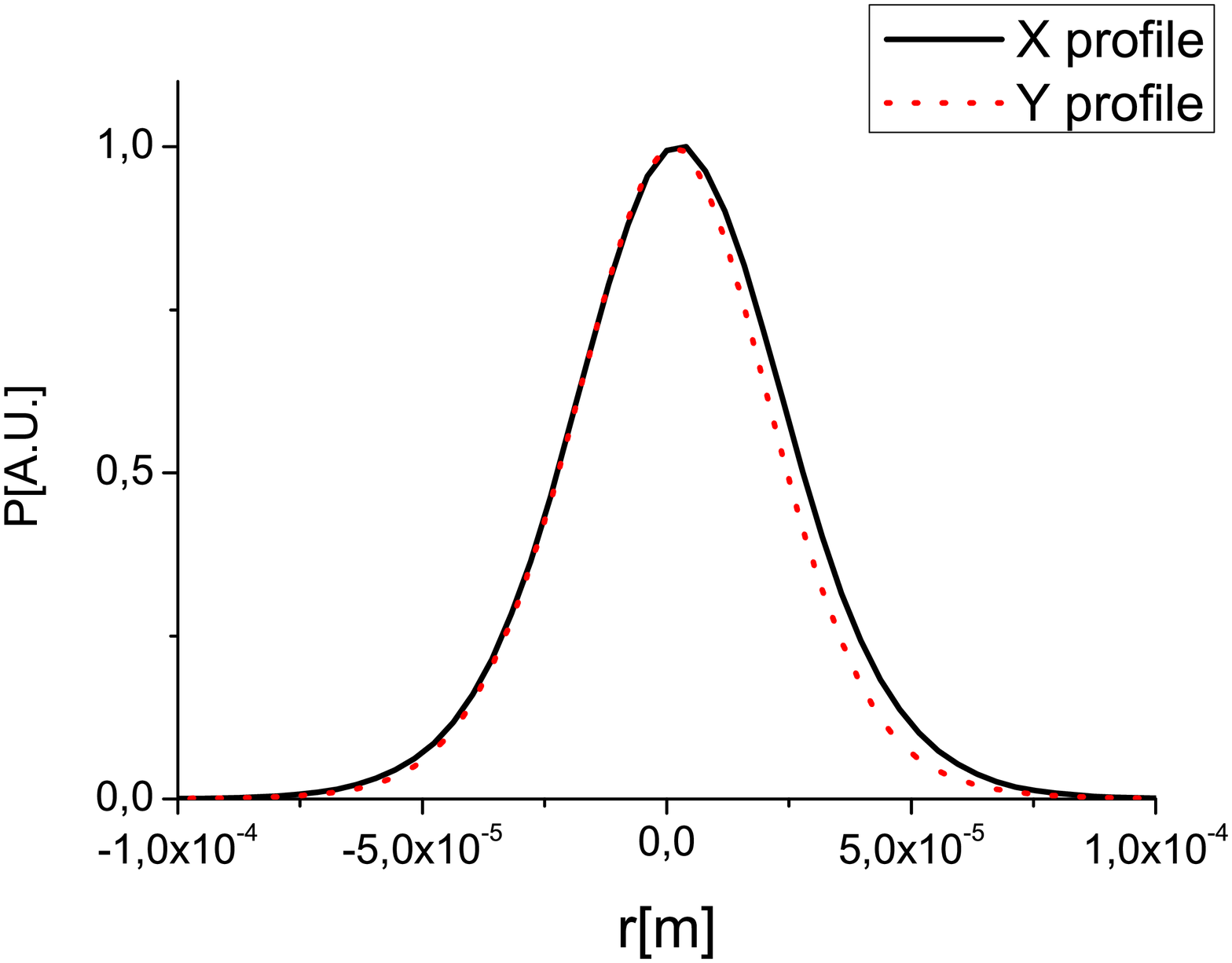}
\includegraphics[width=0.5\textwidth]{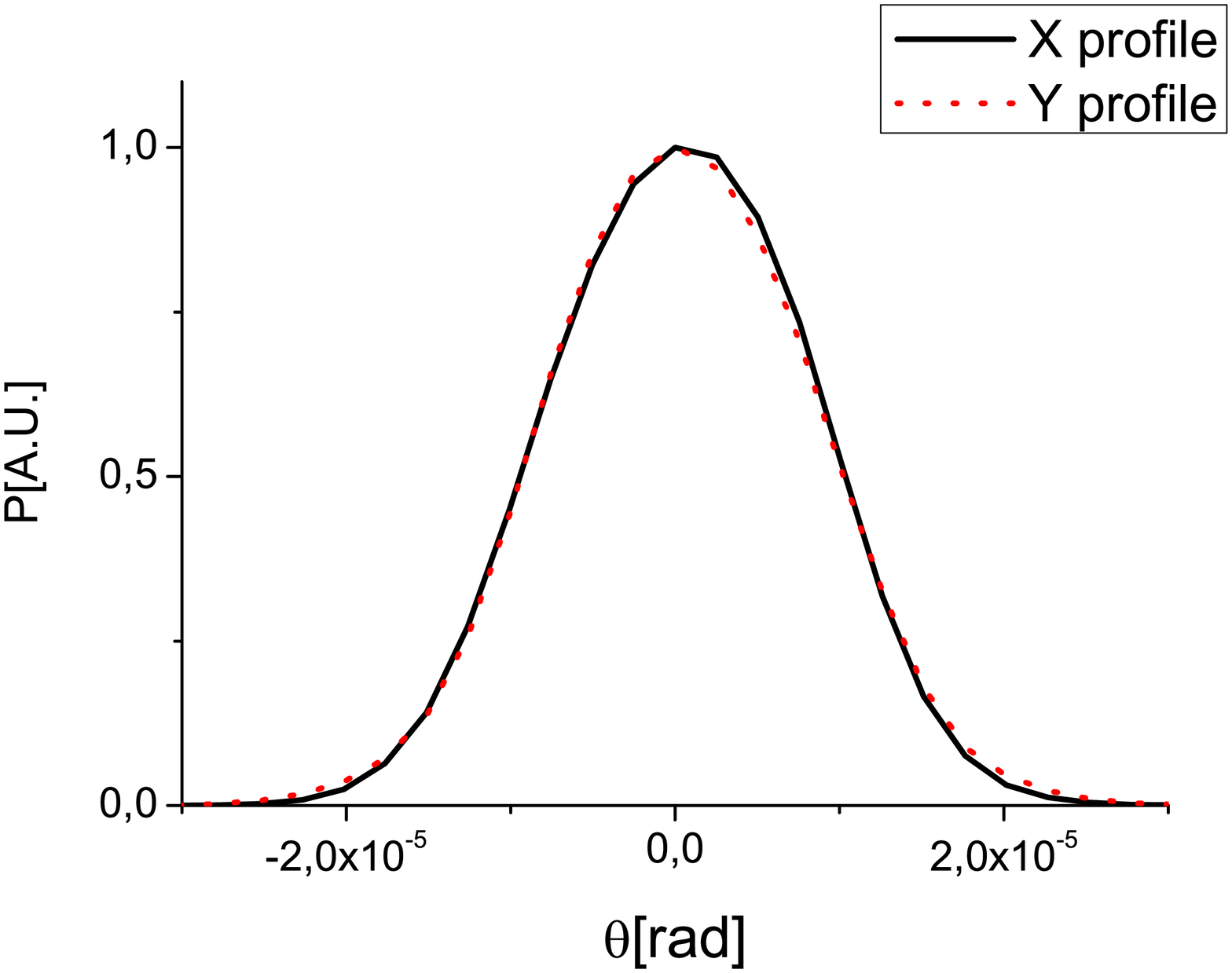}
\caption{Left plot: Transverse plot of the X-ray radiation pulse
energy distribution after the first SASE undulator (5 cells). Right
plot: Angular plot of the X-ray radiation pulse energy distribution
after the first SASE undulator (5 cells).} \label{TrAndis}
\end{figure}
\begin{figure}
\includegraphics[width=1.0\textwidth]{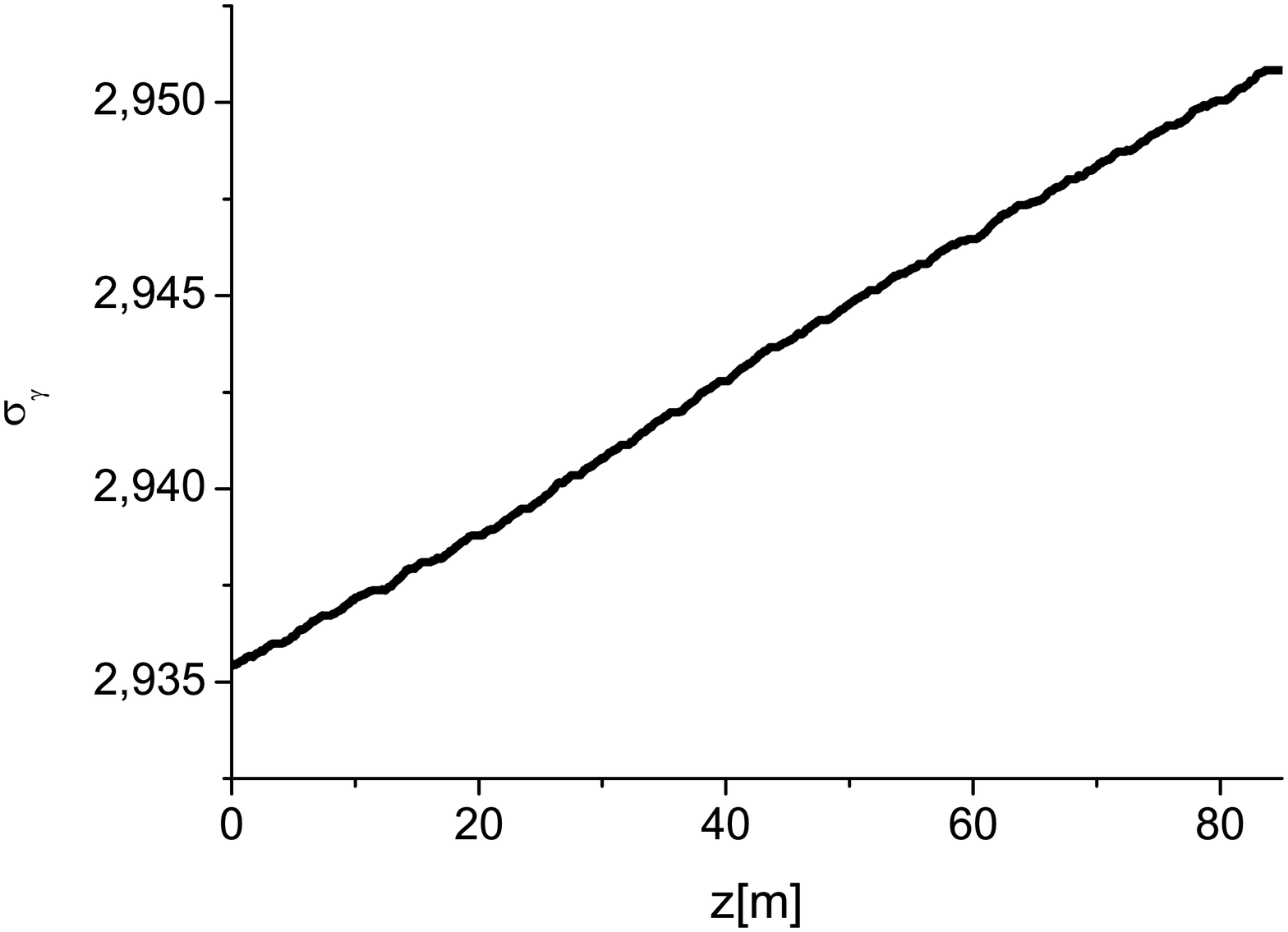}
\caption{The increase of energy spread due to the influence of
quantum fluctuations in the baseline undulator is negligible.}
\label{sigE}
\end{figure}
First, the baseline SASE undulator output was simulated. The result,
in terms of power and spectrum, is shown in Fig. \ref{SASE}, while
the angular distribution of the radiation is shown in Fig.
\ref{TrAndis}. In order to obtain Fig. \ref{TrAndis}, we first
calculated the intensity distribution along the bunch, so that in
the left plot we present the energy density as a function of the
transverse coordinates $x$ or $y$, as if it was measured by an
integrating photodetector. A two-dimensional Fourier transform of
the data finally yields angular X-ray radiation pulse energy
distribution. The $x$ and $y$ cuts are shown on the right plot.

The influence of quantum fluctuations in the baseline undulator was
also studied. Only the five last cells were used, but one needs to
account for the fact that beam passed, before the last five cells,
through many detuned cells. Fig. \ref{sigE} shows that such
influence is negligible.

\begin{figure}[tb]
\includegraphics[width=0.5\textwidth]{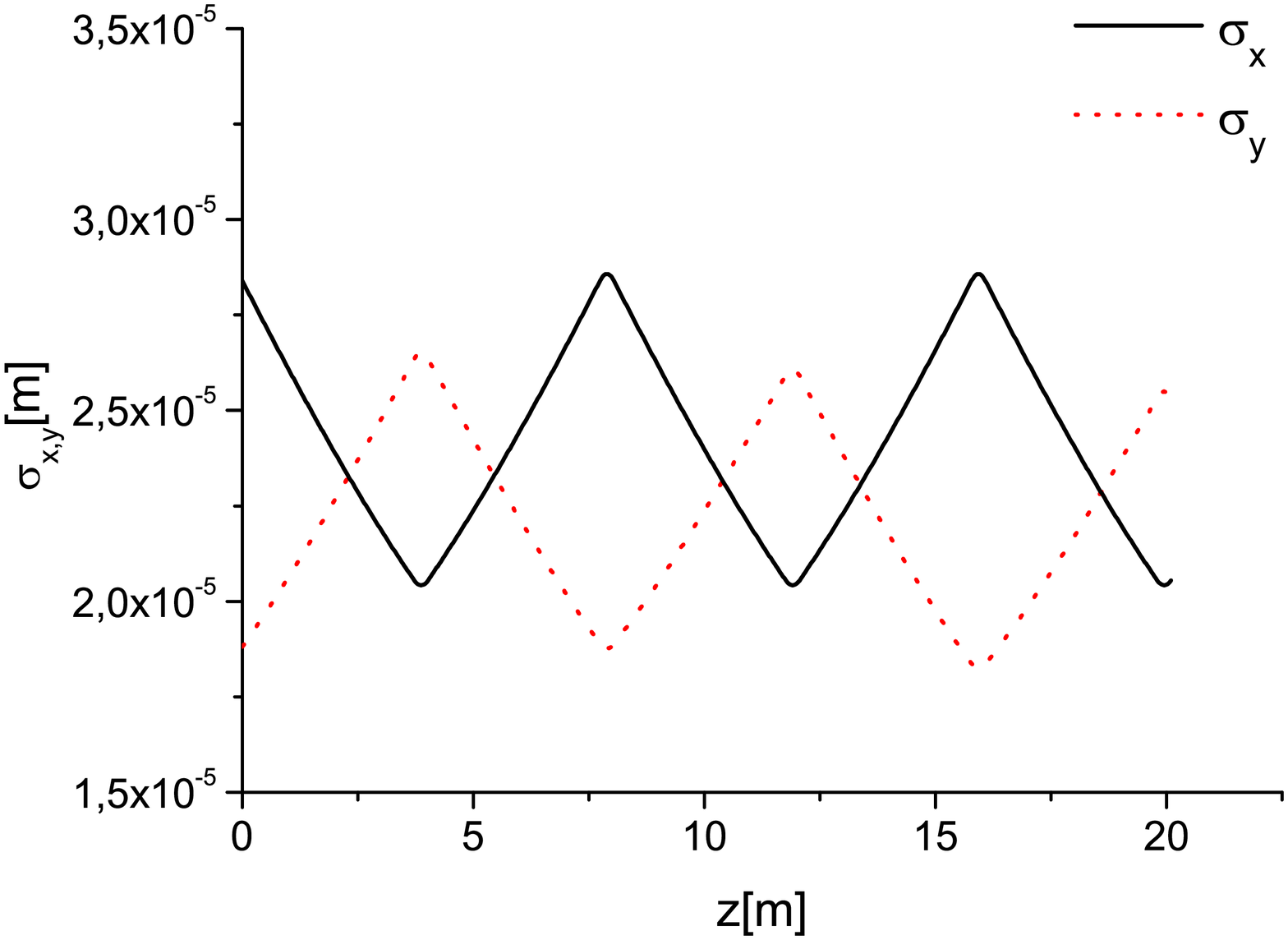}
\includegraphics[width=0.5\textwidth]{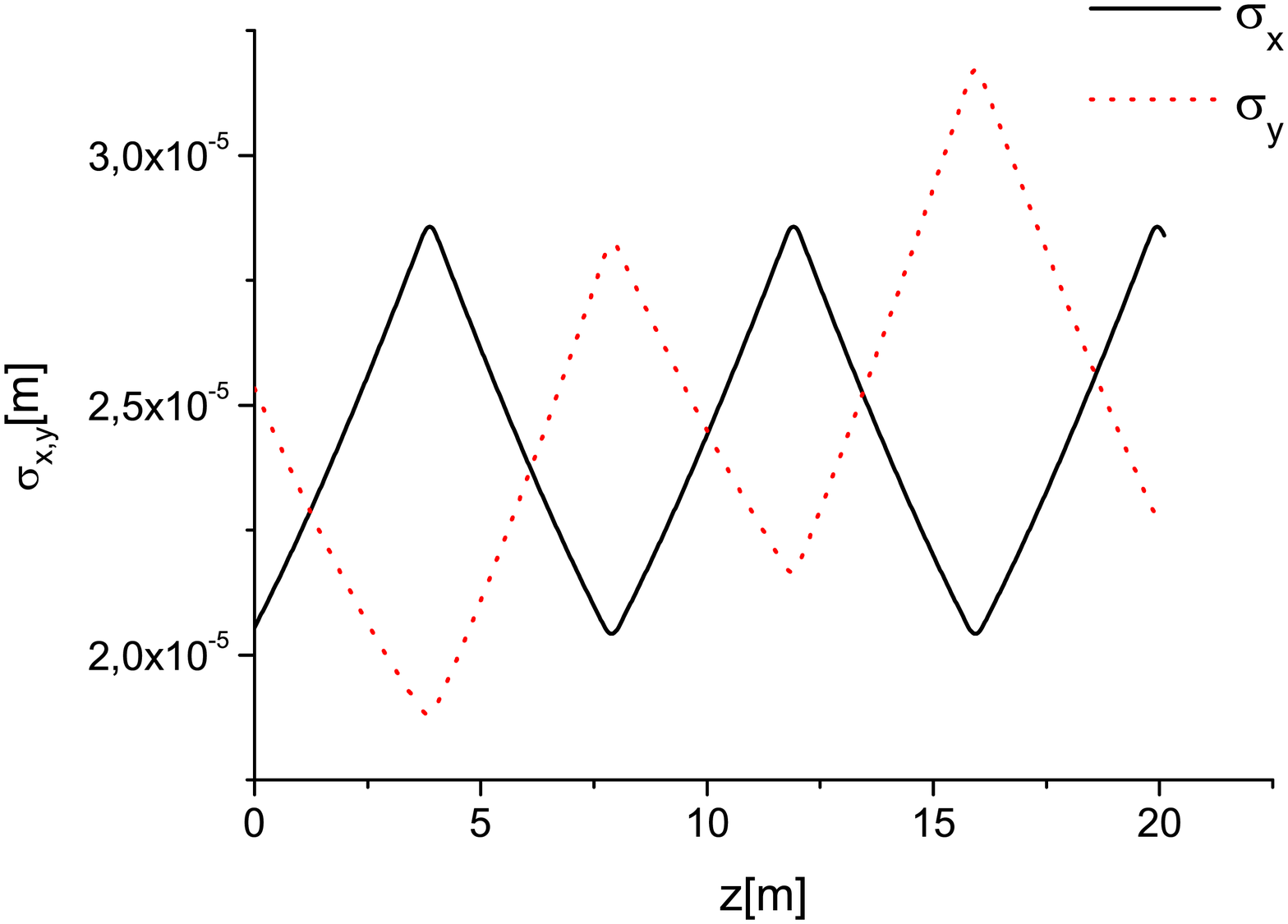}
\caption{Evolution of the rms horizontal and vertical beam size as a
function of the distance along the setup calculated through Genesis.
Left plot: along the baseline undulator. Right plot: along the
straight section following the baseline undulator.} \label{SIGXY}
\end{figure}
\begin{figure}[tb]
\includegraphics[width=1.0\textwidth]{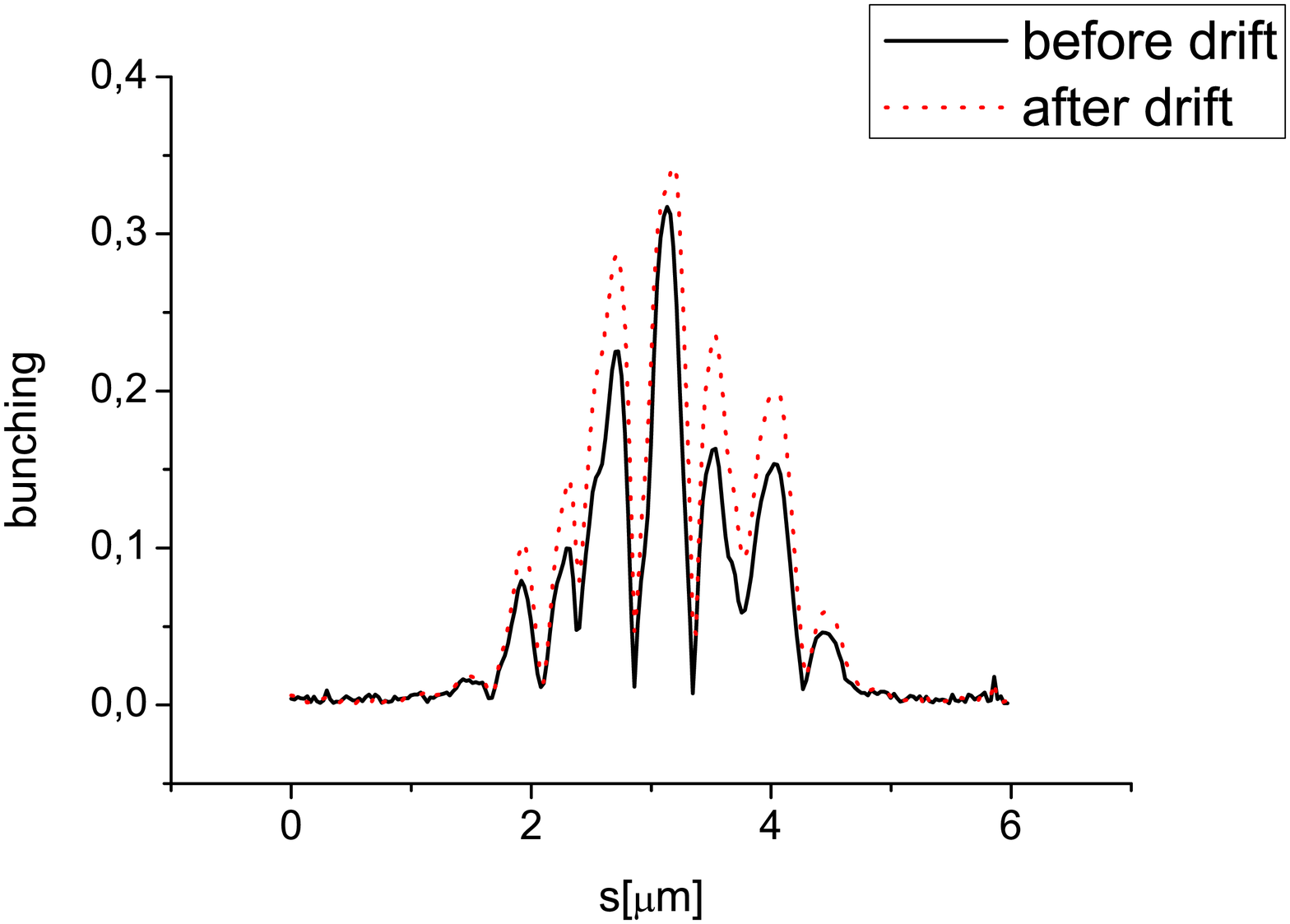}
\includegraphics[width=1.0\textwidth]{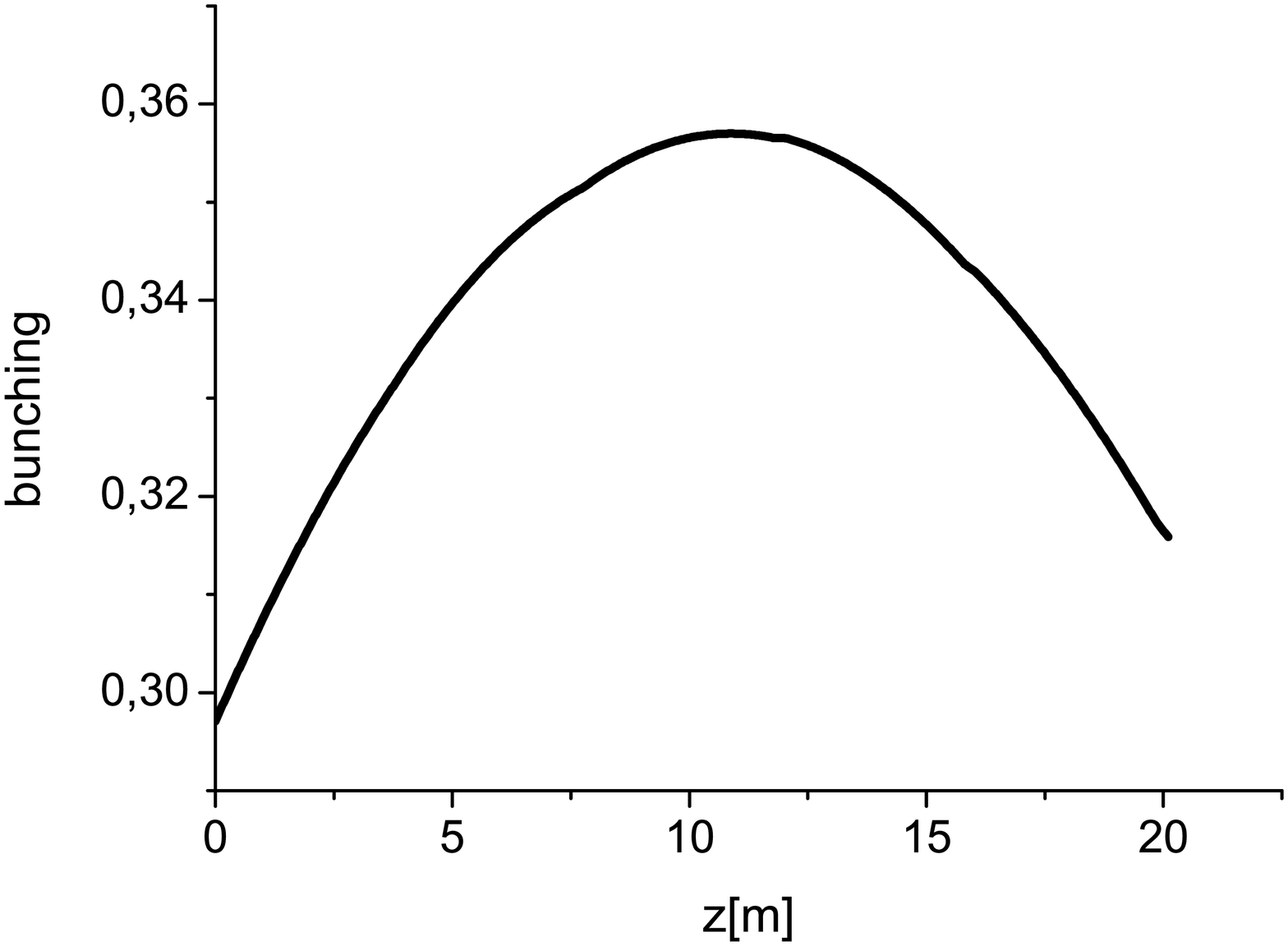}
\caption{Upper plot: comparison between the bunching before and
after the drift for a particular FEL run. Lower plot: evolution of
the bunching in the central slice along the straight section for the
same run.} \label{bunching}
\end{figure}

The GENESIS particle file was downloaded at the exit of the baseline
undulator. For simulating the straight section in GENSIS, we used
the same 5-cells undulator structure as for the baseline undulator,
but we changed the undulator parameter to $K = 0.1$ This choice
allows one to have, with an accuracy of a fraction of percent, the
same momentum compaction factor as in free space. Then the electron
beam current was set to zero, and the undulator focusing was
switched off (although for $K=0.1$ the undulator focusing effects
are negligible). The GENESIS particle file was used as an input for
the propagation of the bunch along the $20$ m-long FODO lattice. The
average betatron function is assumed to be $\beta = 10$ m. GENESIS
automatically accounts for momentum compaction factor and betatron
motion effects on the evolution of the microbunched beam. We tested
the correctness of GENESIS simulations concerning the betatron
motion effects in reference \cite{OUBE}. The bunching before and
after the straight section drift is shown in Fig. \ref{bunching},
upper plot, as a function of the longitudinal coordinate inside the
electron bunch, for a particular run. The evolution of the middle
slice bunching along the straight section for the same run is shown
in Fig. \ref{bunching}, lower plot. Within the straight section, the
bunching increases up to some maximum and then starts to decrease
due to the longitudinal velocity spread originating from the beam
energy spread and, additionally, from the betatron motion.

The evolution of the electron beam size in the horizontal and in the
vertical direction inside the baseline undulator and in the straight
section are shown in Fig. \ref{SIGXY}. Inspection of Fig.
\ref{SIGXY}, right plot, shows a little mismatching in the vertical
direction $y$, which is not present in the left plot. This is due to
the fact that in baseline undulator the electron beam was matched
accounting for the undulator focusing properties. However, the
mismatch was accounted for, because we used the particle file at the
exit of straight section as input file for GENESIS simulations of
the APPLE II output, meaning that the particle file was downloaded
again at the end of the propagation through the straight section.
Note that each cell begins with an undulator, and finishes with a
quadrupole. Therefore we downloaded the particle file immediately
after the first quadrupole related with propagation inside the APPLE
II undulator. This guarantees correct propagation along the APPLE II
undulator section, which is $4.85$ m long. The output files are
downloaded immediately after the  APPLE II undulator.

\begin{figure}
\includegraphics[width=0.5\textwidth]{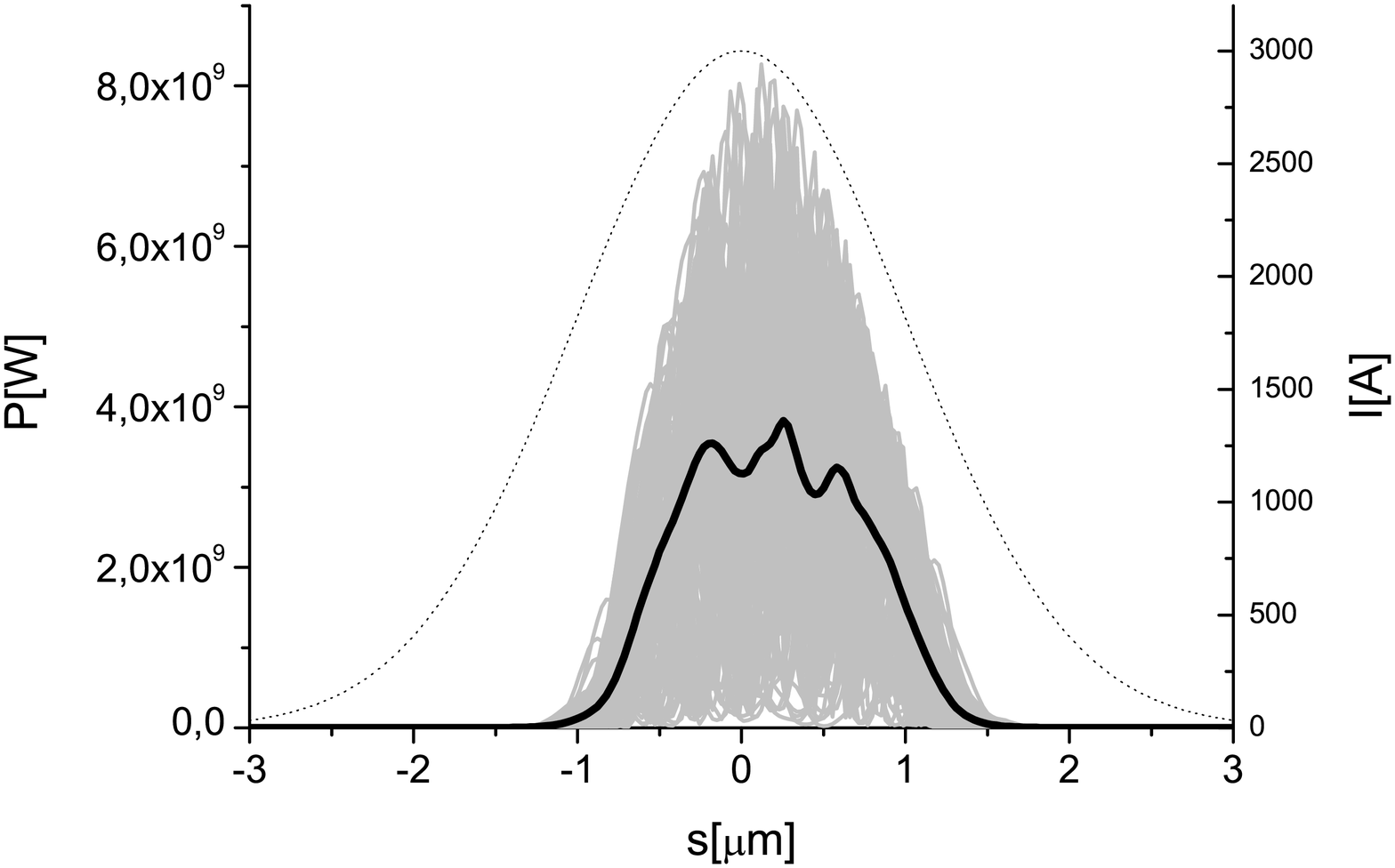}
\includegraphics[width=0.5\textwidth]{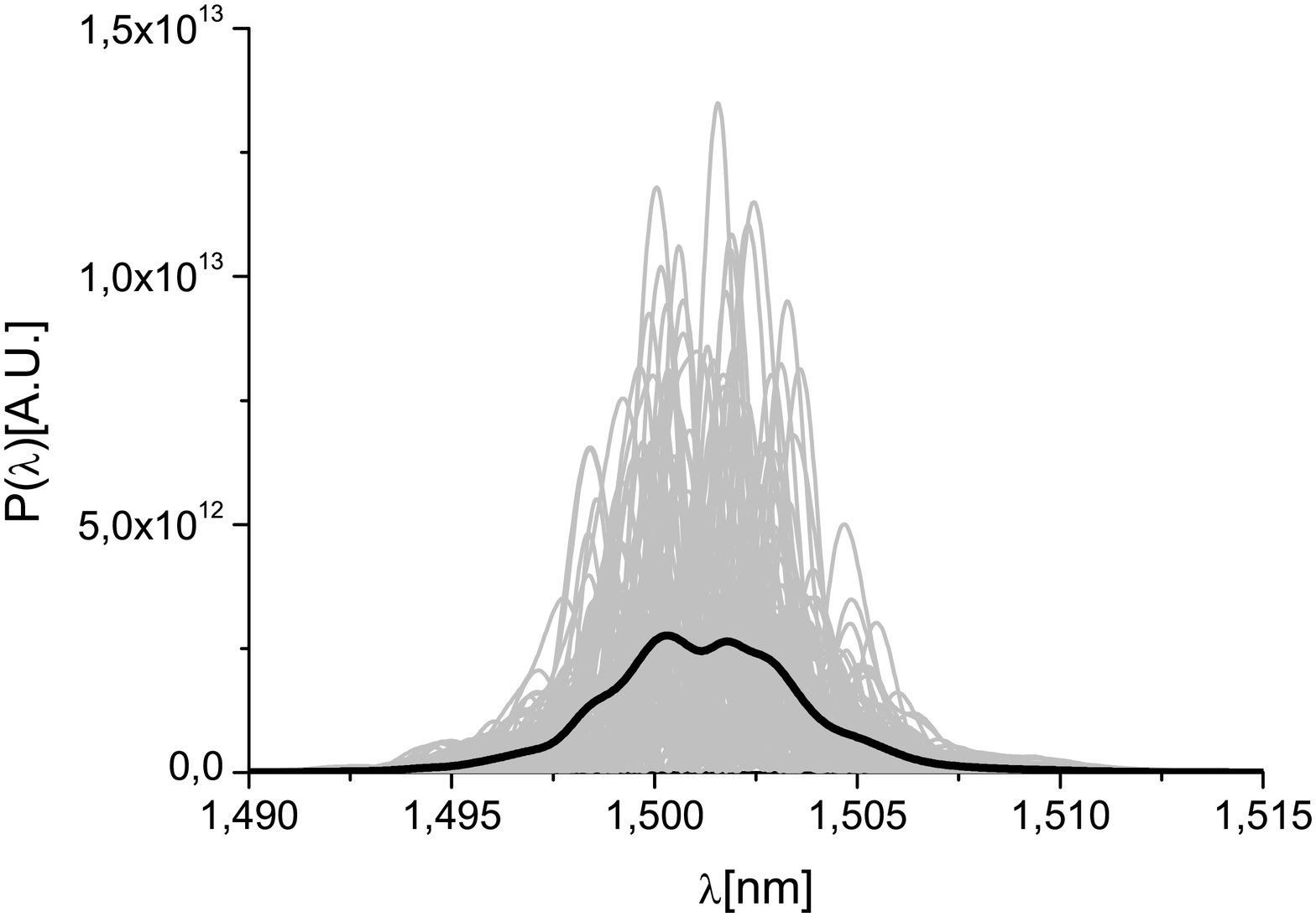}
\caption{Left plot: output power distribution from the APPLE II
undulator. The dotted line refers to the original electron bunch
profile. Right plot: output spectrum. Grey lines refer to single
shot realizations, the black line refers to the average over a
hundred realizations. } \label{output}
\end{figure}
\begin{figure}
\includegraphics[width=0.5\textwidth]{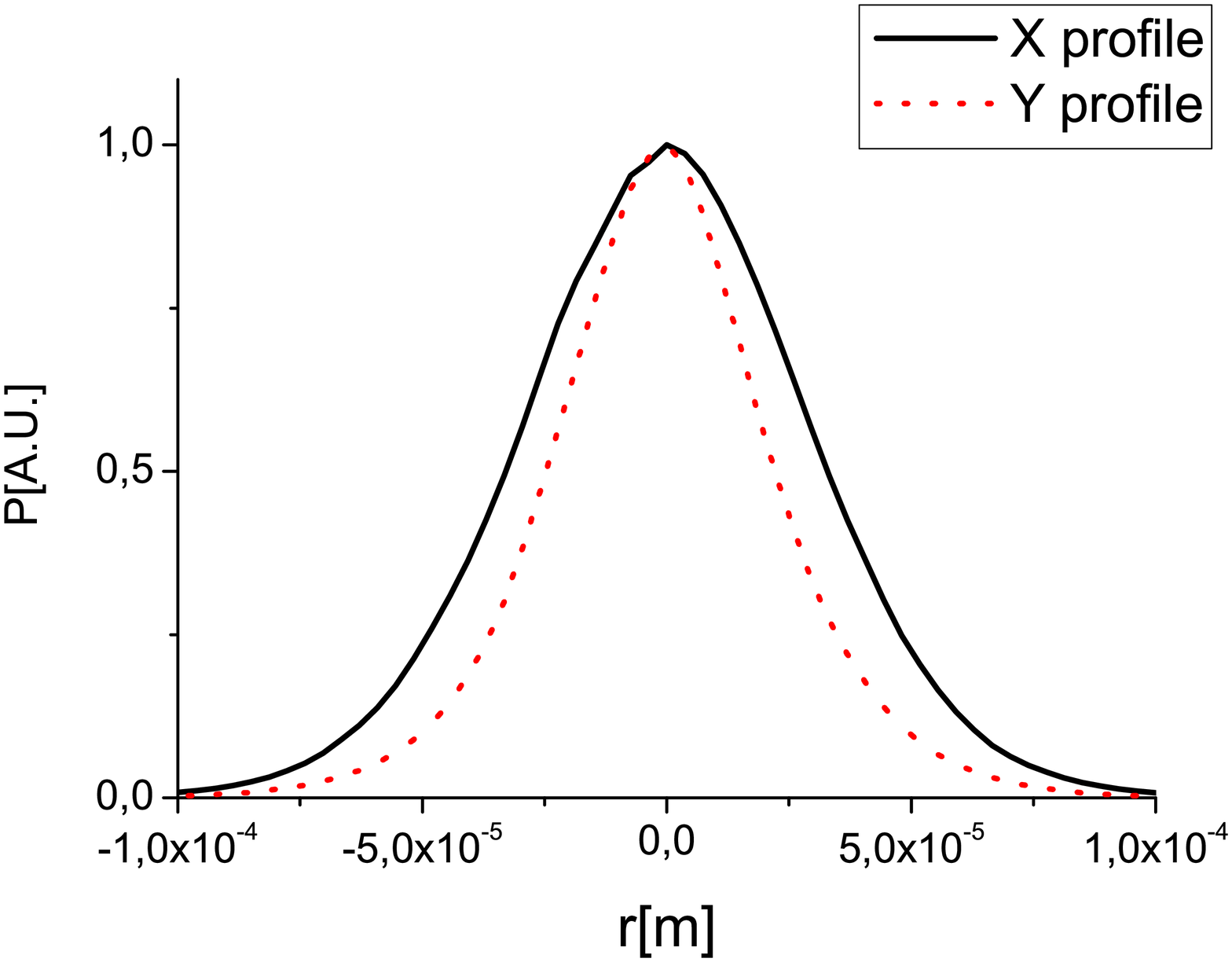}
\includegraphics[width=0.5\textwidth]{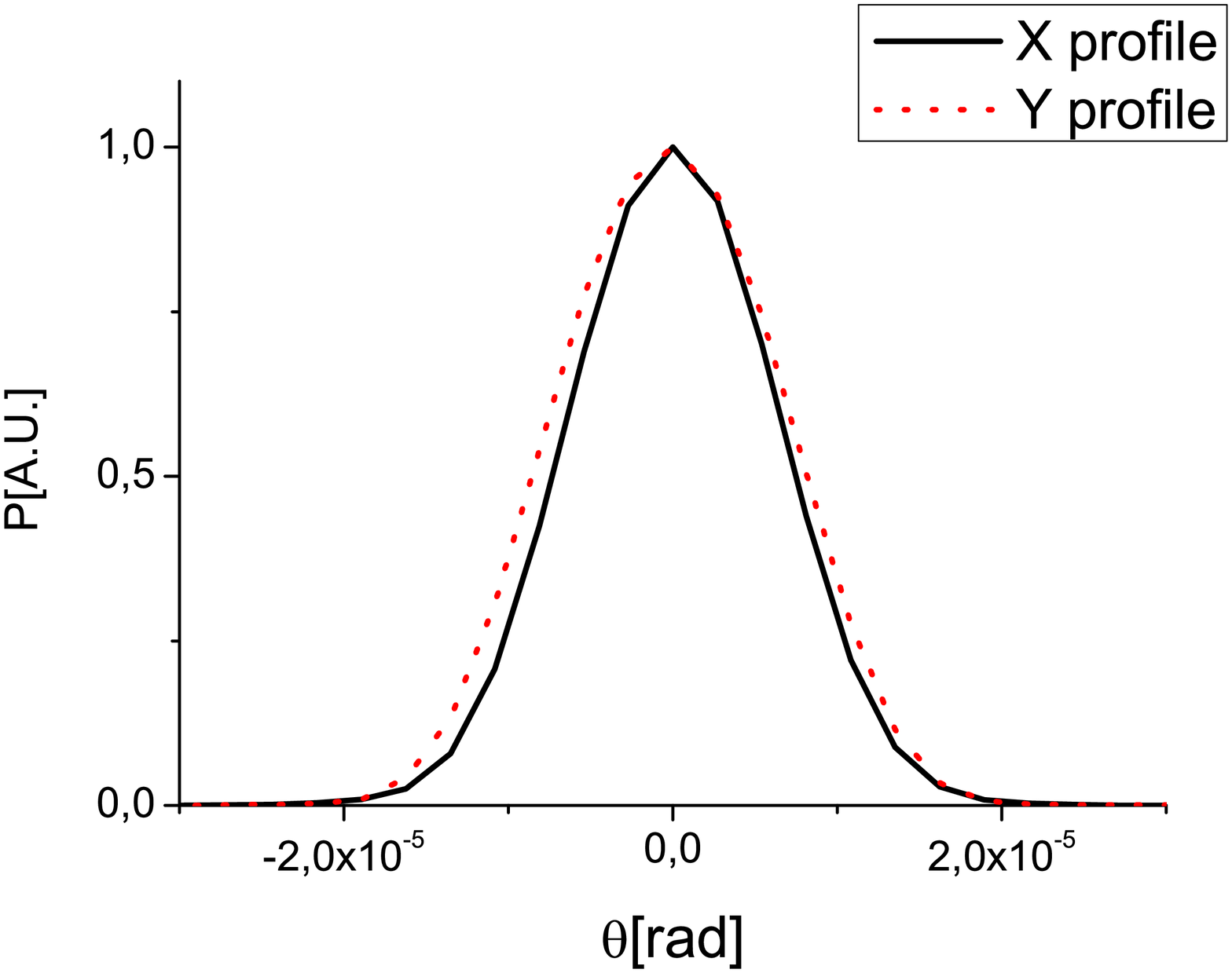}
\caption{Left plot: Transverse plot of the X-ray radiation pulse
energy distribution after the APPLE II undulator (5 cells). Right
plot: Angular plot of the X-ray radiation pulse energy distribution
after after the APPLE II undulator (5 cells).} \label{TrAndis_O}
\end{figure}
The final output from our setup is shown in Fig. \ref{output}, left
plot, in terms of power and in Fig. \ref{output}, right plot, in
terms of spectrum. In order to optimize the APPLE II output, the
fifth section of the upstream planar undulator was detuned of
$\Delta \omega/\omega = -0.02$ by changing the undulator parameter
$K$ from the number corresponding to exact resonance at $1.5$ nm at
$4.3$ GeV. In this way a few-GW output power figure is granted,
which is about the input level reported in Fig. \ref{SASE}. The
transverse distribution of the radiation is shown in Fig.
\ref{TrAndis_O} in terms of transverse coordinates (left plot) and
angles (right plot). From the analysis of Fig. \ref{TrAndis} one
finds an angular size of $20~\mu$rad FWHM. As a result, after $25$ m
propagation, the transverse size of the SASE radiation is about
$0.5$ mm FWHM, to be compared with the APPLE II radiation spot size,
which is just $60~\mu$m. The SASE radiation spot size is, therefore,
about ten times larger than the APPLE II radiation spot size. We
assume, conservatively, the same energy level in both pulses. A slit
system letting through the FWHM of the APPLE II radiation would let
pass a relative contribution of linearized radiation of about
$(60/500)^2 = 0.014$, yielding a degree of circular polarization in
excess of $98 \%$.

\section{\label{sec:conc} Conclusions}

The LCLS baseline does not offer the possibility of polarization
control. The output radiation is simply linearly polarized.
Implementation of polarization control at LCLS baseline is a
challenging problem, subject to many constraints including the
request of a low cost, little available of time to implement setup
changes, and guarantee of safe return to the baseline mode of
operation. It is clear that the lowest-risk strategy for the
implementation of polarization control at the LCLS baseline involves
adding an APPLE II-type undulator at the end of the LCLS baseline
undulator and exploiting the microbunching of the baseline planar
undulator. Detailed experience is available in synchrotron radiation
laboratories concerning the manufacturing of a $5$ m-long APPLE II
undulator. However, the choice of short radiator leads to background
suppression problems. In fact, the linearly-polarized radiation from
the baseline undulator should be separated from the
variably-polarized output from the APPLE II undulator. The driving
idea of our proposal is that the background radiation can be
suppressed by spatial filtering. This operation consists in letting
radiation and electron beam through slits immediately behind the
APPLE II undulator, which is placed immediately behind the whole
($33$ cells) baseline undulator. The estimated cost is low enough to
consider adding this scheme to the LCLS baseline in a two-years
period.

\section{Acknowledgements}

We are grateful to Massimo Altarelli, Reinhard Brinkmann, Serguei
Molodtsov and Edgar Weckert for their support and their interest
during the compilation of this work.

\end{document}